\documentclass[12pt]{article}
\usepackage{latexsym}
\begin{document}
\title{\bf Multi-Scale Perturbation Analysis in Hydrodynamics of the Superfluid Turbulence.
Derivation of the Dresner Equation.}
\author{Sergey K. Nemirovskii$^*$ and  Sergey A. Ponomarenko$^{**}$ \\
$^*$Institute for Thermophysics, Prosp. Lavrentyeva, 1, \\
Novosibirsk, 630090, Russia; E-mail: nemir@itp.nsc.ru \\
$^{**}$Department of Physics and Astronomy, University of
Rochester,\\ Rochester, NY 14627 USA }
 \date{}
 \maketitle

\begin{abstract}
The Hydrodynamics of Superfluid Turbulence (HST) describes the flows (or
counterflows) of HeII in the presence of a chaotic set of vortex filaments, so
called superfluid turbulence. The HST equations govern both a slow variation
of the hydrodynamic variables due to dissipation related to the vortex
tangle and fast processes of the first and second sound propagation. This
circumstance prevents an effective numerical simulations of the problems of
unsteady heat transfer in HeII. By virtue of a pertinent multi-scale
perturbation analysis we show how one can eliminate the fast processes to
derive the evolution equation for the slow processes only. We then
demonstrate that the long-term evolution of a transient heat load of
moderate intensity obeys the nonlinear heat conductivity equation, often
referred to as the Dresner equation. We also compare our approach against
the Dresner phenomenological derivation and establish a range of validity of
the latter.\bigskip\\{\bf Keywords: HeII, Superfluid Turbulence, Unsteady
Heat Transfer}\\
\end{abstract}
\begin{description}
\item[\large Introduction and scientific background]
\end{description}
As it is known, upon exceeding some critical value of the counterflow
velocity in case of the motion induced by the heat load applied to one end
of a channel, a chaotic set (or tangle) of quantized vortex filaments
emerges in superfluid helium. The presence of the vortex tangle dramatically
changes all the hydrodynamic properties of HeII in comparison with those
prescribed by usual two-fluid hydrodynamics (See for details book by
Donnelly \cite{Don} and review article by
Tough \cite{Tough}). Due to a small value of the critical velocity such
situation takes place almost in any experiment on studying hydrodynamic and
heat transfer phenomena in HeII. Consequently, any flow or counterflow of
HeII should be investigated with the help the set of hydrodynamics equations coupled with
Vinen equation for the vortex line density
dynamics (see \cite{Vinen}) . This set of equations known as the
Hydrodynamics of Superfluid Turbulence (HST) have been obtained in papers of
Nemirovskii and Lebedev \cite{NL}, Geurst \cite{Geurst} and Yamada et. al.
\cite{Yamada} within the scope of different approaches. The Hydrodynamics of
Superfluid Turbulence as well as the methods for its investigation are
outlined in the review paper by Nemirovskii and Fiszdon \cite{NF}.
The set of HST equations is extremelly cumbersome, therefore,
there is no wonder that to achieve quantitative results in
virtually any case of practical interest one is bound to turn to
numerical methods. Yet, numerical simulation of nonstationary
flows of HeII faces one serious obstacle. The point here is that
the set of HST equations that unifies the hydrodynamics equations
of superfluid helium and the Vinen equation for the vortex line
density evolution is initially of hyperbolic type. As a result, a
slow variation of hydrodynamic variables due to dissipation is
accompanied by the fast processes related to propagation (and
possible reflections from the boundaries of the channel) of the
first and second sounds. If one is interested in the slow
evolution of the temperature, velocities and the vortex line
density only, particular details of sound propagation and manifold
reflections come out completely irrelevant, yet requiring rather
extensive numerical resources. Indeed, a typical time of transient
processes is of order of seconds, whereas a typical mesh size in
numerical investigations of acoustic phenomena should be of order
of at least microseconds to account for the forming of the shock
fronts. {\it Thus, it seems attractive to try to get rid of the
fast modes by a pure analytical procedure}. In the present paper
we realize effective separation of the slow from the fast modes
using multi-scale (in the case multi-time) asymptotic perturbation
techniques (See, for example, book by Nayfeh \cite{Nayfeh}. The
main idea here is to introduce several time scales, each of which
responsible for the description of a corresponding process. As an
example we recall, perhaps, the simplest possible case of the
overdamped harmonic oscillator. Given a certain relation between
the frequency and the damping decrement, sinusoidal oscillations
die away shortly, whereas slow exponential relaxation may take
quite a while to die out. Therefore, the long-term dynamic
behavior changes drastically. The same is true for our case:
dimensional analysis of the HST equations shows that there is a
dimensionless criterion, Strouhal number $Sh$ that is the ratio of
the counterflow decrement to inverse time that takes the heat
pulse to cross the channel. Depending on the input parameters of
the problem the number $Sh$ can be either large or small. The
latter circumstance allows us to come up with an effective
procedure for separation of the fast from the slow modes. The
paper is organized as follows. The second section is devoted to
dimensionless analysis of the set of HST equations. In the third
section we carry out the multi-scale asymptotic perturbation
procedure for the case of heat load of moderate intensity, of
order of a few $W/cm^2$ i.e. for $Sh>>1$. We then establish that
the slow processes are governed by the nonlinear heat conductivity
equation similar to that derived phenomenologically by Dresner
\cite{Dresner}. We also discuss the criteria of validity of the
Dresner equation and the range of its applicability. In conclusion
we resume the obtained results.
\begin{description}
\item[\large Dimensionless equations of HST]
\end{description}
As it was noted above the full set of the HST equations is very
complicated and in general, does not permit any analytical
investigation at all. Having in mind to study processes related to
heat transfer in HeII, we propose the following
simplifications.\\(i) First, we restrict ourselves to the case of
counterflow of HeII which occurs whenever one applies a heat load
to a reservoir filled with superfluid helium.\\(ii) Second, we
study only quasi-one-dimensional cases, i.e. either pure
one-dimensional ({d=1}), or cylindrical ({d=2}), or spherical
({d=3}) geometries. \\(iii) Next, we neglect by the nonlinear
convective terms on the left-hand sides of the equations of
motion. Usually these terms come in play in the cases of very
large heat fluxes (when no separation of the "fast" and "slow"
mode is
possible at all) and are responsible for the forming of the shock fronts. \\
(iv) Finally, we drop the dissipation function which is very small in the
cases to be studied.The above assumptions correspond to the situations
brought about in the overwhelming majority of experiments on transient heat
transfer.
    To describe the counterflow of HeII it is convenient to choose the following
set of variables: dimensionless velocity of the normal component $
V_n^{\prime }$ , dimensionless temperature $T^{\prime }$ and dimensionless
square root of the vortex line density (VLD) $G$ ($G=\sqrt{{\cal L}/{\cal
L_\infty }}$ , where ${\cal L_\infty \ }$ is equilibrium with respect to the normal velocity value of VLD).
Under the assumptions listed above the set of equation of HST (see
for details e.g. review article of Nemirovskii and Fiszdon \cite{NF}) is
reduced to the following form:


\begin{equation}  \label{V}
\frac{\partial V_n^{\prime }}{\partial t^{\prime }}\;+\;\frac{\partial
T^{\prime }}{\partial x^{\prime }}\;=\;-\;Sh\;G^2V_n^{\prime }\;
\end{equation}
\begin{equation}  \label{T}
\frac{\partial T^{\prime }}{\partial t^{\prime }}\;+\;\frac 1{x^{\prime
}{}^{d-1}}\;\frac \partial {\partial x^{\prime }}{x^{\prime
}{}^{d-1}V_n^{\prime }}\;=0,
\end{equation}
\begin{equation}  \label{G}
\frac{\partial G}{\partial t^{\prime }}\;+\;M_L\frac{\partial G}{\partial
x^{\prime }}\;=\;-\;\frac{Vi\;Sh}2\;(G^2V_n^{\prime }\;-\;G^3),
\end{equation}
with the following dimensionless variables:
\begin{equation}  \label{DIM}
t\;=\;\frac LU\;t^{\prime },\qquad x\;=\;Lx^{\prime },\qquad
V_n\;=\;V_{n0}\;V_n^{\prime },\qquad T\;=\;\frac{\sigma V_{n0}}{{\sigma }_TU}
T^{\prime },\qquad {\cal L}\;=\;{\cal L}_\infty G^2.
\end{equation}
Here $L$ is a length of the channel, $U$ is the velocity of the second
sound, $V_{n0}$ is a characteristic value of the normal velocity at the
boundary $V_{n0}=q/\rho \sigma T$ (here $q$ is a heat flux, which is
actually a function of time, $\sigma $ is entropy per unit mass and $\sigma_{T}$ is
a derivative of entropy with respect to the temperature).


At this stage it is worth making a few remarks concerning the set of
equations (\ref{V})-(\ref{G}). If one used dimensional variables $V_n,T$ and
${\cal L}$ instead of their dimensionless counterparts, one would easily
recognize the equations widely used in studying the dynamics of intense heat
pulses. Nevertheless a somewhat awkward use of variable $G$ in lieu of the vortex
line density is necessary to develop the multi-scale perturbation procedure.


It is easy to see from set (\ref{V})-(\ref{G}) that the dynamics of the
hydrodynamic variables is specified by several dimensionless criteria,
namely the Vinen parameter $Vi$, the Strouhal parameter $Sh$ and the Mach
number $M_L$. They are defined as:
\begin{equation}  \label{par}
Sh\;=\;\frac L{U{\tau }_d},\qquad M_L=\frac{V_L}U,\qquad Vi=\;\frac \alpha
{A(T) {\rho }_s{\rho }_n}\;(\frac \alpha \beta )
\end{equation}


Here we have introduced the following notations: $\alpha $ and $\beta $, the
parameters of the Vinen equation; $A(T)$ the Gorter-Mellink constant; $V_{L}$
is a drift velocity of the vortex tangle;  ${\tau
}_d$, a characteristic time of attenuation of the second sound due to the
vortex tangle introduced by Vinen (see \cite{Vinen}):
\begin{equation}  \label{t_d}
{\tau }_d=\frac{\rho _s^2}{A\rho ^3V_{n0}^2}.
\end{equation}

Numerical estimation of parameters (\ref{par}) shows that within the
temperature interval 1.4 -- 2.1 K parameter $Vi$ changes from 0.2 to 1.5. It
does not depend on applied heat flux $q$ (or on quantity $V_{n0{\rm {\ }}} $
). The Mach number, although dependent on the heat flux, remains very small.
As far as parameter $Sh$ is concerned, it strongly depends on the heat flux
and can be either large or small compared to unity:
\begin{equation}  \label{Sh small}
Sh\;<<\;1,\;\;\;{\rm {for\;\;\;small\;\;\;heat\;\;\;flux}}
\end{equation}
\begin{equation}  \label{Sh large}
Sh>>1,\;\;\;{\rm {\ for\;\;\;large\;\;\;heat\;\;\;flux}}
\end{equation}


Thus, the main parameter, drastically affecting the behavior of the system,
is the Strouhal parameter $Sh$ which is nothing else than the ratio of the
counterflow damping decrement due to an interaction with the vortex tangle
to the inverse time of the flight of the heat pulse. Conditions (\ref
{Sh
small})-( \ref{Sh large}) can be used for effective separation of the
processes with characteristic time $t_0\sim L/U$ (second sound behavior)
from those with characteristic time $t_1\sim (L/U)\;/Sh$ (evolution due to
counterflow-vortex tangle mutual friction).


Having in mind to investigate the cases of practical interest, we estimate $
Sh$\ numbers for heat load of moderate intensity $q=1\;\div \;10\;\;W/cm^2$
and channel sizes $L=10\div 10^2cm$. Using the definitions (\ref{par}), (\ref
{t_d}) of the Strouhal number as well as thermodynamical parameters we
obtain that in the temperature region $T=1.4\div 2.1$K the Strouhal number
is of order of $10^2-10^5$, i.e. much greater than unity. Therefore, we can
restrict ourselves to studying the cases of large $Sh$\ numbers.


\begin{description}
\item[\large Multi-time scale method]
\end{description}


In this section we carry out elimination of the fast processes. As mentioned
above we will concentrate on the case of large Strouhal parameter $Sh>>1$.
Put in another way, attenuation of the second sound is assumed so strong
that the convective regime of heat transfer quickly gives way to another
regime, reminiscent of nonlinear heat conduction in usual newtonian fluids.
Formally this leads to the degeneration of the initially hyperbolic
equations into a parabolic equation describing the slow evolution of the
velocity and temperature fields. To demonstrate this transition explicitly we
invoke a pertinent multi-scale asymptotic perturbation theory (see e.g.Nayfeh
\cite{Nayfeh}). Following this method, we introduce different time scales.


\begin{equation}  \label{t}
t_0=t^{\prime};\qquad \;t_1^{\prime}=\epsilon t^{\prime};
\qquad t_2^{\prime}=\epsilon ^2t^{\prime} {\rm {\ \qquad
......., } }
\end{equation}
where $\epsilon =1/Sh<<1$. We look for a solution to the set of the
HST equations (\ref{V})-(\ref{G}) in the form of an asymptotic series


\begin{equation}  \label{VTL}
\begin{array}{c}
V_n^{\prime }=V_0^{\prime }(x^{\prime },t_0^{\prime},t_1^{\prime},
t_2^{\prime})+\epsilon V_1^{\prime
}(x^{\prime },t_0^{\prime},t_1^{\prime},t_2^{\prime})+
\epsilon ^2V_2^{\prime }(x^{\prime},t_0^{\prime},t_1^{\prime},t_2^{\prime})+.... \\
T^{\prime }=T_0^{\prime }(x^{\prime },t_0^{\prime},t_1^{\prime},t_2{\prime})
+\epsilon T_1^{\prime}
(x^{\prime },t_0^{\prime},t_1^{\prime},t_2^{\prime})+\epsilon ^2T_2^{\prime }
(x^{\prime },t_0^{\prime},t_1^{\prime},t_2^{\prime})+....
\\
G=G_0(x^{\prime },t_0^{\prime},t_1^{\prime},t_2^{\prime})+\epsilon G_1
(x^{\prime},t_0^{\prime},t_1^{\prime},t_2^{\prime})+\epsilon ^2G_2
(x^{\prime },t_0^{\prime},t_1^{\prime},t_2^{\prime})+....
\end{array}
\end{equation}


In accordance with multi-scale
perturbation analysis (see \cite{Nayfeh})
the coefficients in the series for dimensionless normal velocity $
V_n^{\prime }$, dimensionless temperature $T^{\prime }$ and qauntity $G$ are
supposed to have an order of smallness $O(\epsilon ^0)$, i.e. to be of order
of unity. The simple chain rule follows from relations (\ref{t}) - (\ref{VTL}):
\begin{equation}
\frac \partial {\partial t^{\prime }}=\frac \partial {\partial t_0^{\prime}}
+\epsilon \frac \partial {\partial t_1^{\prime}}+
\epsilon ^2\frac \partial {\partial t_2^{\prime}}
\label{d/dt}
\end{equation}


The next step in study of the slow evolution of the heat pulse consists in
substituting the multi-time scale series (\ref{t})-(\ref{VTL}) into
equations of HST (\ref{V} )-(\ref{G}). Gathering terms of the same order of
magnitude with respect to $\varepsilon$ we come up with a chain of equations
leading to divergent (secular) solutions. Canceling step by step these
secularities we then obtain a hierarchy of equations of different orders in
parameter $\varepsilon $, governing different stages of the evolution of the
fields of temperature, counterflow velocity and the square root of the vortex
line density.


Let us develop the outlined scheme in detail. Excluding the temperature
variable $T^{\prime }$ from the set of equations (\ref{V})-(\ref{G} ) we
rewrite it in the as:
\begin{equation}  \label{G1}
G^2\;V_n^{\prime }\;-\;G^3\;=\; - \frac 2{Vi}\epsilon \left( \frac{\partial G}{
\partial t^{\prime }}\;+\;M_L\frac{\partial G}{\partial x^{\prime }}\right) ,
\end{equation}
\begin{equation}  \label{V1}
\frac \partial {\partial t^{\prime }}\left( G^2\;V_n^{\prime }\right)
\;=\;\epsilon \left[ \frac \partial {\partial x^{\prime}}\;\left( \frac 1{x^{\prime
}{}^{d-1}}\;\frac \partial {\partial x^{\prime }}{x^{\prime }{}^{d-1}}V{\
_n^{\prime }}\right) \;-\;\frac{\partial ^2V{_n^{\prime }}}{\partial
t^{\prime }{}^2}\right] .
\end{equation}


Implementation of the above procedure depends on a particular statement of
the problem under consideration. To be specific, we assume that at the
moment a heater is switched on the thermal front starts out propagating into
undisturbed bulk of helium which is, in fact, the case in most experiments.
Such statement, however, brings up some apparent paradox. Indeed, substituting
series (\ref{VTL}) for $G$ and $V$ and using rule (\ref{d/dt}) we find
out that to zero order in $\varepsilon $, quantity $G$ is equal to $
V_0^{\prime }(x^{\prime },t_0^{\prime},t_1^{\prime},t_2^{\prime})$. Put in
another way, it follows that
VLD ${\cal L}(t)$ immediately takes its equilibrium value with respect to
the normal velocity. But that is wrong as there actually is some finite time
$\tau _V\;$ of development of the VLD (see \cite{Vinen}). Hence, we seem to
end up in quite a contradictory situation. To find a way out, let us compare
the characteristic time of the vortex tangle development $\tau _V\;$ against
the characteristic time of propagation of heat pulse $L/U$. As it is well
known (see \cite{Vinen} ), the former is determinated by relation:
\begin{equation}
\tau _V\;=\;a(T)\;q^{-3/2}.  \label{Tv}
\end{equation}
Taking into account a scatter of data it follows
\begin{equation}
q^{3/2}L\;\sim \;40\;\div \;140\left[ \frac{W^{3/2}}{cm^2}\right] .
\label{QL}
\end{equation}
It is seen from (15) that for heat pulses of moderate amplitude $1\;\left[
W/cm^2\right] \;\leq \;q\;\leq \;10\;\left[ W/cm^2\right] $
propagating in channels of length $L\;\sim 10\div 10^2cm$ , which are
typical of experiments on transient heat transfer in HeII, the quantities $
\tau _V\;$ and $L/U$ are of the same order of magnitude. It implies that
for the time of order $(L/U)$, typical of the slow processes, the vortex
line density has enough time to adjust to the velocity field, i.e. to take
its equilibrium (with respect to the relative velocity) value
${\cal L}_\infty \;=\;\frac{\rho V_n^2}{{\rho }_s}(\frac \alpha \beta )^2$.
 In other
words, if we start our scheme not at $t=0$ when there are no vortices in the
bulk of helium but at time $t=\tau _V$ when the VLD has adjusted to its
equilibrium value, we can then proceed without any discrepancies. At times
smaller than $\tau _V$ when $G$ is small relations (\ref{G1})-( \ref{V1})
can be rewritten as


\begin{equation}
\left[ \frac \partial {\partial x^{\prime }}\frac 1{x^{\prime
}{}^{d-1}}\;\frac \partial {\partial x^{\prime }}{\ x^{\prime }{}^{d-1}}
\;-\; \frac{\partial ^2}{\partial t^{\prime }{}^2}\right] \;V_n^{\prime }\;=\;0
\label{Ve1}
\end{equation}


\begin{equation}
G^2\;V^{\prime }\;=\;\frac{\partial G_0}{\partial t^{\prime }}\;+\;M_L\frac{
\partial G}{\partial x^{\prime }}  \label{Ge1}
\end{equation}
The meaning of relations (\ref{Ve1}) and (\ref{Ge1}) is obvious. For times smaller
than  $L/U$ the heat pulse propagates according to wave-like equation
because of the absence of vortices ( excluding extremely small background
value). The vortex line density, being very small in comparison with its
equilibrium value ${\cal L}_\infty $, changes mainly by first generating
term of the Vinen equation\footnote{
To see what would exactly happen at times smaller than $\tau _V$ one had to
resolve a fairly difficult problem of propagation of the heat pulse which creates the vortices on its way and interacts with the very vortices. The corresponding
investigations are described in the review article \cite{NF}}.


Let us now proceed with the general scheme. The following step in studying the slow evolution
of the heat pulse consists in substituting multi-scale variables (\ref{t})-(
\ref{VTL}) into equations ( \ref{V1})-(\ref{G1}). Gathering terms of the
same order, we wind up with the chain of equations:


\underline{To order $\epsilon ^0$.}
\begin{equation}
\begin{array}{c}
V_0^{\prime} = V_0^{\prime}(x^{\prime },t_1^{\prime},t_2^{\prime}) \\
G_0=V_0^{\prime}.
\end{array}
\label{zero}
\end{equation}


\underline{To order $\epsilon ^1$}.
\begin{equation}
\begin{array}{c}
V_1^{\prime} \;-\;G_1\;=\;\frac{2M_L}{Vi}\frac 1{V_0^{\prime 2}}\frac{\partial V_0^{\prime}}{\partial
x^{\prime }}, \\
V_0^{\prime 2}\frac \partial {\partial t_0^{\prime }}\left( V_1^{\prime}\;+\;2G_1\right)
\;=\;-\frac \partial {\partial t_1^{\prime }}G_0^3\;+\;\frac \partial
{\partial x^{\prime }}\left( \frac 1{x^{\prime d-1}}\frac \partial {\partial
x^{\prime }}(x^{\prime d-1}\;V_0^{\prime})\right) .
\end{array}
\label{first}
\end{equation}
Using (\ref{zero}) the second equation of set (\ref{first}) can be
integrated over $t_0^{\prime }$ to yield:
\begin{equation}
\begin{array}{c}
3V_0^{\prime 2} V_1^{\prime}\;=\;f(x^{\prime },t_1^{\prime},t_2^{\prime})\;+\;\frac{4M_L}{Vi}\frac 1{V_0^2}\frac{
\partial V_0^{\prime}}{\partial x^{\prime }}\;+ \\
\;t_0^{\prime}\left[ \frac \partial {\partial x^{\prime }}\frac 1{x^{\prime
}{}^{d-1}}\frac \partial {\partial x^{\prime }}(x^{\prime d-1}\;V_0^{\prime})-\frac
\partial {\partial t_1^{\prime }}G_0^3\;\right] .
\end{array}
\label{first1}
\end{equation}
It is seen that for $t_0^{\prime }\longrightarrow \infty $ the quantity $
V_1^{\prime}\longrightarrow \infty $ as well. In order to cancel this divergence we
have to require the quantity inside the brackets to be equal to zero. This
procedure is called the cancellation of divergencies in secular terms in an
asymptotic series. Having accomplished this procedure, we figure out that to
the first order in $\epsilon ^1$ the evolution of the heat pulse obeys the
set of equations below:
\begin{equation}
V_1^{\prime}=V_1^{\prime}(x^{\prime },t_1^{\prime},t_2^{\prime}),  \label{firstA}
\end{equation}


\begin{equation}
V_1^{\prime}\;-\;G_1\;=\;\frac{2M_L}{Vi}\frac 1{V_0^{\prime 2}}\frac{\partial V_0^{\prime}}{\partial
x^{\prime }},  \label{firstB}
\end{equation}


\begin{equation}
\frac \partial {\partial t_1^{\prime }}V_0^3\;=\;\frac \partial {\partial
x^{\prime }}\frac 1{x^{\prime d-1}}\frac \partial {\partial x^{\prime
}}(x^{\prime d-1}\;V_0),  \label{firstC}
\end{equation}


Thus we have derived a set of equations which describes slow variations of
hydrodynamic variables ( with the values of parameters
discussed above).  It can then be inferred from (\ref{zero}) and (\ref{firstC}) that the leading part
of the normal velocity depends only on the slow time and is governed by the {\it
parabolic-type evolution equation.} This is the main result of the paper. It holds
true provided that all the necessary creteria for application of the outlined
perturbation procedure are met (see discussion above). Later we shall establish
the range of validity of our approach. At this point, however, we complete the calculation.
To this end let us study the next, $\epsilon ^2$ order. We follow exactly the same
procedure as above, cancelling divergences one step at a time. Leaving out
simple but somewhat tedious calculations we write down the final result:


\underline{To order $\epsilon ^2$}.
\begin{equation}
\begin{array}{c}
V_0^{\prime}\;=\;V_0^{\prime}(x^{\prime },t_1^{\prime}) \\
V_1^{\prime}\;=\;V_1^{\prime}(x^{\prime },t_1^{\prime}) \\
G_1=G_1(x^{\prime },t_1^{\prime}) \\
G_0\;=\;V_0^{\prime} \\
V_1^{\prime}\;-\;G_1\;=\;\frac{2M_L}{Vi}\frac 1{V_0^{\prime 2}}\frac{\partial V_0^{\prime}}{\partial
x^{\prime }} \\
\frac \partial {\partial t_1^{\prime }}V_0^{\prime 3}\;=\;\frac \partial {\partial
x^{\prime }}\frac 1{x^{\prime }{}^{d-1}}\frac \partial {\partial x^{\prime
}}(x^{\prime d-1}\;V_0^{\prime}) \\
3\;\frac \partial {\partial t_1^{\prime }}(V_0^{\prime 2}V_1^{\prime})\;=\;\frac \partial
{\partial x^{\prime }}\frac 1{x^{\prime }{}^{d-1}}\frac \partial {\partial
x^{\prime }}(x^{\prime d-1}\;V_1^{\prime})\;+\;\frac{4M_L}{Vi}\;\frac 1{V_0^{\prime 2}}\frac{
\partial V_0^{\prime}}{\partial x^{\prime }}.
\end{array}
\label{Second}
\end{equation}


It is plain that the last equation accounts for a small correction to the
velocity field and hence is of little interest to us.


\begin{description}
\item[\large The Dresner nonlinear heat conductivity equation]
\end{description}


Let us consider the first $\epsilon ^1$ order in greater detail. With the
goal to compare our results to the well-known Dresner heat conductivity
equation(see e.g.\cite{Dresner}) we consider here only the simplest geometry $
(d=1)$. Combining equation [21] with the results
of the previous section we arrive at the set of dimensionless equations for the normal
velocity, and the temperature:
\begin{equation}
\frac{\partial V_0^{\prime 3}}{\partial t_1^{\prime }}\;=\;\frac{\partial ^2V_0^{\prime}}{
\partial x^{\prime }{}^2},  \label{Drv}
\end{equation}
\begin{equation}
\frac{\partial T^{\prime }}{\partial t^{\prime }}\;+\;\frac{\partial V_0^{\prime}}{
\partial x^{\prime }}=0,  \label{DrT}
\end{equation}
Here $V_0^{\prime},T^{\prime},$ are zero terms in series (\ref{VTL}) for the dimensional normal
velocity and the temperature. We recall that to this order in $\epsilon $ the
vortex line density takes its equilibrium value with respect to normal
velocity , e.i. ${\cal {L}}{}=V_0^{\prime 2}.$  Next, excluding quantity $V_0^{\prime}$ from
the set of equations written above we obtain the relation:
\begin{equation}
\frac{\partial T^{\prime }}{\partial t_1^{\prime }}\;=\;\epsilon
^{-1/3}\frac \partial {\partial x^{\prime }}\left( \frac{\partial T^{\prime }
}{\partial x^{\prime }}\right) ^{1/3}  \label{DR}
\end{equation}
Relation (\ref{DR}) formally coincides with the widely used nonlinear
heat-conductivity equation derived by Dresner \cite{Dresner}. In this
connection it is worth discussing the method used by Dresner. He started
with the Gorter-Mellink relation which in our notation transforms to:
\begin{equation}
\frac{\partial T^{\prime }}{\partial x^{\prime }}\;=\;-Sh\;V^{\prime }{}_n^3
\label{dT/dx}
\end{equation}
Recalling that heat flux $q$ is related to normal velocity $V_n$ via $
q\;=\;STV_n$, he concluded that the heat flux was proportional to the cube
root of the temperature gradient.


\[
q\propto \left( \frac{\partial T^{\prime }}{\partial x^{\prime }}\right)
^{1/3}.
\]
Furthermore, using the energy conservation law,
\[
\frac{\partial S}{\partial t}+{\rm {\ }}\frac{\partial SV_n}{\partial x}=0.
\]


Dresner derived an equation similar to equation (\ref{DR}). His
approach, however, is not self-consistent from point of view of the full set
the HST equations. Indeed the Gorter-Mellink relation (\ref{dT/dx} )
corresponds to a steady-state and its direct use for a nonstationary
case is not valid. The correct equation should be obtained from equations
(1), in which quantity $G$ is replaced by its equilibrium value $V^{\prime
}{}_n$.
\[
\frac{\partial V^{\prime }{}_n}{\partial t^{\prime}}+\frac{\partial T^{\prime }}{
\partial x^{\prime }}\;=\;-Sh\;V^{\prime }{}_n^3
\]


Throwing out this term implies that $V_n^{\prime }=V_n^{\prime }(x)$ which,
in turn, means that the temperature is also a function of $x$ only and does
not change in time. Accordingly $T^{\prime }$ connected to $V_n^{\prime }$
via Gorter-Mellink relation does not change in time.


In our approach $\partial V_0^{\prime }/\partial t_1^{\prime }\neq 0$ and
there is no contradiction. Our careful consideration explicitly demonstrates
that the regime of the nonlinear heat-conductivity equation takes place only
for ''slow'' time $(\sim L/U)$ and validity of the overall procedure requires meeting the
criteria dwelt upon above. The proposed allows also to determine a range of
validity of the Dresner
equation (\ref{DR}). One restriction is that slow regime takes place
at a time scale of order $(L/U)$ therefore a parabolic type heat-transfer
equation
works at times greater than $\tau_d$. The latter  then provides the lower
limit of the appplicability of
the evolution equation derived above, equation (23).


On the other hand, the obvious upper limit is furnished by the boiling time
calculated in \cite{NKB}.


To conclude this chapter we would like to point out the region of
fulfillment of condition (\ref{Sh large}) in terms of the heat flux and the
size of the channel. Using the definition of the Strouhal number (\ref{par})
and thermodynamical parameters, we obtain that in the temperature region $
T=1.4\div 2.1$K condition (\ref{Sh large}) is equivalent to the following
one:
\begin{equation}
q^2L\;>>\;0.2\;\div \;0.6[\frac{W^2}{cm^3}].  \label{Sh'}
\end{equation}
Thus for heat load of moderate intensity $q=1\;\div \;10\;\;W/cm^2$ and for
sizes of the setup $L=10\div 10^2cm$ relation (\ref{Sh large}) and
consequently (\ref{Sh'} ) are valid to good accuracy.


\begin{description}
\item[\large Conclusion]
\end{description}


We have outlined the procedure of separation of the fast from the slow
processes in the equations of Hydrodynamics of Superfluid Turbulence. As an
illustration of the developed procedure, we studied the slow stage of the
evolution of the transient heat load of moderate intensity. It has been
shown how the wave-like behaviour of the heat pulse described by initially
hyperbolic equations gives way to the parabolic-type non-linear heat
conductivity equation. The latter formally coincides with the one
derived earlier by Dresner on the phenomenological grounds. Unlike the
approach advocated by Dressner, our consideration is not only self-consitent but
also allows to determine a range of validity of the elimination procedure.
\bigskip\\{\bf {\large Acknowledgements}}\bigskip\\
This work was partially supported by grant N 03-02-16179 of the
Russian Foundation of Fundamental Research and by grant N
2001-0618 from INTAS.



\end{document}